\documentclass[useAMS]{mn2e}
\usepackage{graphicx,amssymb}
\title[Cosmic Ray Acceleration by Relativistic Shocks: Limits and Estimates]
{Cosmic Ray Acceleration by Relativistic Shocks: Limits and Estimates}
\author[AR Bell, AT Araudo, JH Matthews, KM Blundell  ]
{A.R. Bell$^{1}$\thanks{E-mail:Tony.Bell@physics.ox.ac.uk},
A.T. Araudo$^{2}$, J.H. Matthews$^{3}$ and K.M. Blundell$^{3}$\\
$^{1}$University of Oxford, Clarendon Laboratory, Parks Road, 
Oxford OX1 3PU, UK\\
$^{2}$Astronomical Institute of the Czech Academy of Sciences, Bocni II 1401, CZ-14100 Prague, Czech Republic\\
$^{3}$University of Oxford, Astrophysics, Keble Road, 
Oxford OX1 3RH, UK\\
}

\begin{document}
\date{}
\pagerange{\pageref{firstpage}--\pageref{lastpage}} \pubyear{2014}
\maketitle
\label{firstpage}
\begin{abstract}
We examine limits to the energy to which cosmic rays can be accelerated by relativistic shocks, showing that acceleration of light ions as high as 100 EeV is unlikely. The implication of our estimates is that if ultra-high energy cosmic rays are accelerated by shocks, then those shocks are probably not relativistic.
\end{abstract}

\section{ INTRODUCTION}
The origin of ultra-high energy cosmic rays (UHECR) is still a mystery.   
A number of cosmic ray (CR) acceleration processes have been proposed.
Here we concentrate on diffusive shock acceleration 
(Axford, Leer \& Skadron 1977, Krimskii 1977, Bell 1978, Blandford \& Ostriker 1978).
General arguments (Hillas 1984; Blandford 2000) indicate that acceleration to $\sim 100\ {\rm EeV}$ is favoured by high shock velocities,  large magnetic fields, large spatial scales and shocks with a high power throughput as may occur in outflows from active galactic nuclei (AGN) and gamma-ray bursts (GRB).
These factors point towards  relativistic shocks as likely accelerators of UHECR,
but here we give reasons why this is not the case.
In this paper we show that, if UHECR are accelerated by shocks, then those shocks are probably not relativistic.

We discuss three effects that limit the maximum energy to which CR can be accelerated by relativistic shocks:
\newline
(i) {\it Steep CR spectrum.}  Relativistic shocks generate CR energy spectra that are steeper than those generated by non-relativistic shocks.  Consequently, there is less energy in the high energy CR component to drive the turbulence and amplify the magnetic field that scatters the high energy CR and determines the rate at which CR are accelerated (Section 3).
\newline
(ii) {\it Small-scale turbulence.} Because CR do not penetrate far upstream of relativistic shocks and because anisotropy in the CR momentum distribution decays rapidly downstream of the shock, the ${\bf j}_{\rm CR} \times {\bf B}$ force (where ${\bf j}_{\rm CR}$ is the CR electric current and $\bf B$ is the magnetic field)  has time to drive turbulence only on scales much smaller than a UHECR Larmor radius and consequently does not effectively scatter UHECR (Section 4).
\newline
(iii){\it Quasi-perpendicular shocks.}  Because relativistic shocks are characteristically quasi-perpendicular, the scattering magnetic field has to be amplified within one UHECR Larmor radius of the shock for effective scattering before UHECR are advected downstream away from the shock (Section 5).

The structure of the paper is that, after two introductory sections, we progressively add in each of these three effects and show how each successively reduces the maximum possible CR energy.  The results are presented in Table 1 and Figure 1, to which we will refer during the course of the paper.  In Section 6  we compare relativistic and non-relativistic shocks.
In Section 7 we discuss the acceleration of heavier nuclei, but in all earlier parts of the paper we consider only protons.
In section 8 we present our conclusions.
Throughout this paper we use SI units except where explicitly stated.


\begin{table*}
\caption{Summary of limits on the maximum CR energy for general $\beta$, $\beta=2.5$ \& $\beta=2.23$.
As plotted in Figure 1,
$\xi _{B} (\beta)=10^{(9 \beta+22.7 )/\beta}$,
$\xi _s (\beta)= 10^{(9 \beta +6.24 )/(\beta -0.8)}$,
$\xi _\perp (\beta) =10^{(9 \beta - 16.8 )/(\beta -2)}$, 
$\xi _{\parallel} (\beta )= 10^{(9 \beta +1.1 )/(\beta -1)}$.
}
\centering 
\begin{tabular}{l c c c}
\hline 
\hline
Energy (eV) & General expression &  $\beta =2.5$ &  $\beta =2.23$   \\ [.5ex] 
\hline
\hline
\multicolumn{4}{l}{Section 3:  Bohm limit with magnetic field in equipartition with the CR energy density at the maximum CR energy} \\
 [2ex] 
$ T_{\rm Bohm}$ ({\rm eq. 8}) & 
$ \xi _B (\beta)\  \Gamma _{\rm sh} ^{(\beta -1)/\beta } n_4^{1/\beta} R_{\rm kpc}^{2/\beta} $ & 
$ 1.2\times 10^{18}  \Gamma _{\rm sh} ^{0.6 }n_4^{0.4}R_{\rm kpc}^{0.8} $ & 
$ 1.5\times 10^{19}\Gamma _{\rm sh} ^{0.55 }n_4^{0.45}R_{\rm kpc}^{0.90} $  
\\
 [2ex] 
$ T_{\rm Bohm}$ ({\rm eq. 10}) & 
$ \xi _B (\beta)\  \Gamma _{\rm sh}^{-(\beta-2)/\beta }  P_{40} ^{1/\beta} $ &
$ 1.2\times 10^{18} \Gamma _{\rm sh} ^{-0.2 } P_{40} ^{0.4}$ & 
$ 1.5\times 10^{19} \Gamma _{\rm sh}^{-0.10} P_{40} ^{0.45} $
\\
 [2ex] 
\hline 
\multicolumn{4}{l}{Section 4:  Magnetic field in equipartition with CR as in Section 3 but with scale-size limited by ${\bf j_{\rm CR} \times {\bf B}}$ displacement} \\
 [2ex] 
$ T_{\rm s}$ ({\rm eq. 14}) & 
$ \xi _s (\beta)    \Gamma _{\rm sh} ^{(5\beta -7)/(5 \beta -4)} n_4 ^{3/(5\beta -4)} R_{\rm kpc}  ^{6/(5\beta -4)} $ & 
$8.1 \times 10^{16}\Gamma _{\rm sh} ^{0.65}n_4 ^{0.35}R_{\rm kpc}  ^{0.71}$ & 
$2.5\times 10^{18} \Gamma _{\rm sh} ^{0.58} n_4 ^{0.42} R_{\rm kpc}  ^{0.84} $
\\
 [2ex] 
\hline 
\multicolumn{4}{l}{Section 5.1:  Perpendicular shock, CR acceleration limited by time needed for magnetic field amplification} \\
 [2ex] 
$ T_{{\rm NRH},\perp}  ({\rm eq. 21}) $ & 
$\xi _{\perp} (\beta)\  \Gamma _{\rm sh} ^{(2\beta-3)/(2 \beta -4)}  n_4^{1/(2 \beta -4)} B_{0,\mu {\rm G}} ^{-1/( \beta -2)}$ & 
$2.5\times 10^{11} \Gamma _{\rm sh} ^{2}  n_4 B_{0,\mu {\rm G}} ^{-2}$ &
$ 1.6\times 10^{14}\Gamma _{\rm sh} ^{3.17} n_4^{2.17}B_{0,\mu {\rm G}} ^{-4.35} $
\\
 [2ex] 
\hline 
\multicolumn{4}{l}{Section 5.2:  Parallel or low-magnetization shock, CR acceleration limited by time needed for magnetic field amplification} \\
 [2ex] 
$
T_{{\rm NRH},\parallel } ({\rm eq. 23}) $ &
$ \xi _{\parallel} (\beta)\    \Gamma _{\rm sh} ^{(2\beta -3)/(2 \beta -2)} n_4^{1/(2 \beta-2)} R_{\rm kpc} ^{1/(\beta -1)}$ & 
$5.4\times 10^{15} \Gamma _{\rm sh} ^{0.67}n_4^{0.33}R_{\rm kpc} ^{0.67}$  & 
$ 1.6\times 10^{17} \Gamma _{\rm sh} ^{0.59} n_4^{0.41} R_{\rm kpc} ^{0.81} $
\\
 [2ex] 
$
T_{{\rm NRH},\parallel } ({\rm eq. 24}) $ &
$ \xi _{\parallel} (\beta)\    \Gamma _{\rm sh}^{(2\beta -4)/(2 \beta -2)} P_{40}^{1/(2\beta -2)}$ & 
$5.4\times 10^{15} \Gamma _{\rm sh}^{0.33}P_{40}^{0.33} $  & 
$ 1.6\times 10^{17} \Gamma _{\rm sh}^{0.19} P_{40}^{0.41} $
\\
 [2ex] 
\hline
\hline
\end{tabular}
\label{table:nonlin}
\end{table*}

\begin{figure}
\includegraphics[angle=0,width=8.3cm]{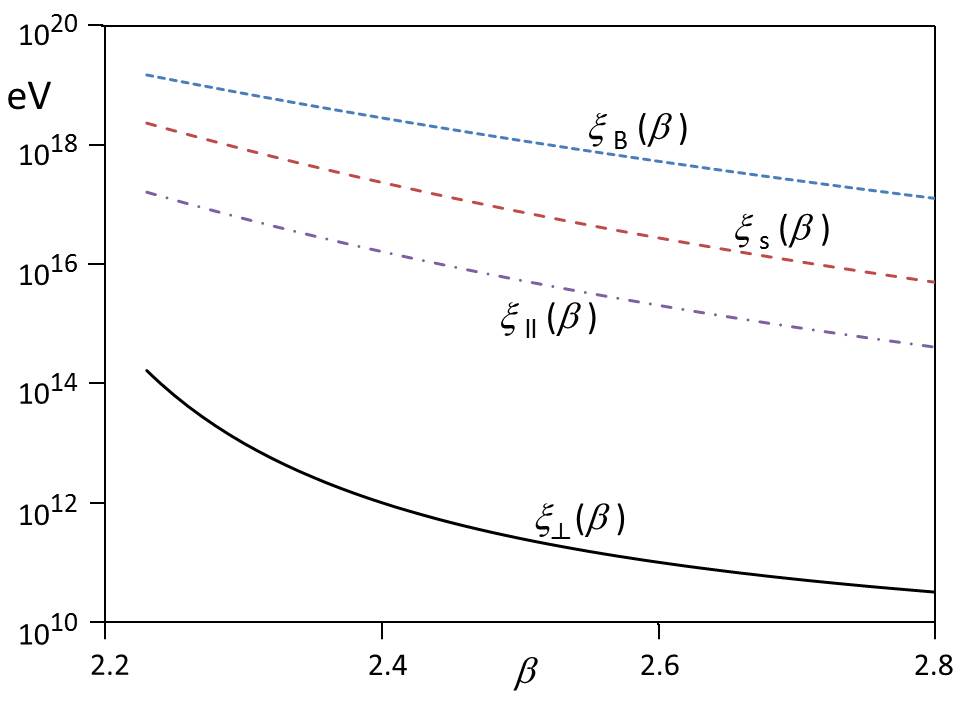}
\caption{
Plots of the scaling factor $\xi (\beta )$ for different limits on the CR energy:
$\xi _{B} (\beta)=10^{(9 \beta+22.7 )/\beta}$ (eq 8 \& 10);
$\xi _s (\beta)= 10^{(9 \beta +6.24 )/(\beta -0.8)} $ (eq 14);
$\xi _\perp (\beta) =10^{(9 \beta - 16.8 )/(\beta -2)}$ (eq 21);
$\xi _{\parallel} (\beta )= 10^{(9 \beta +1.1 )/(\beta -1)}$ (eq 23 \& 24).
$\xi$ corresponds to the maximum energy in eV of CR accelerated by a relativistic shock with our characteristic parameters for shock size $R$ $ (1{\rm kpc})$, plasma
density ($10^{-4}{\rm cm}^{-3}$), jet power ($10^{40}{\rm W}$) and  uniform magnetic field $B_{0,\perp}$ ($1 \mu{\rm G}$).
}
\label{fig:figure1}
\end{figure}

\section{UPPER LIMITS ON THE MAXIMUM CR ENERGY}

As discussed by Hillas (1984) an upper limit to the maximum energy to which a CR 
particle can be accelerated by a shock is $T_{\rm H}=uBR$ where   
$R$ is the size of the accelerating region, $u$  is the characteristic velocity of the bulk thermal plasma, and $T_{\rm H}$ is the particle kinetic energy in eV.  
The Larmor radius of an UHECR 
is orders of magnitude larger than that of thermal particles, so the bulk thermal plasma behaves magnetohydrodynamically (MHD)
with a local electric field ${\bf E}=-{\bf u}\times {\bf B}$.  
Consequently $uB$ is the maximum electric field available and $uBR$ corresponds to moving the particle a distance $R$ through an electric field $uB$.  
The Hillas energy can be written as
$$
T_{\rm H}=\left (\frac{B}{\mu {\rm G}}\right )  \left (\frac{u}{c} \right ) \left (\frac{R}{\rm kpc}\right ) \ 0.9\  {\rm EeV}
\  .
\eqno(1)
$$
An alternative interpretation of this expression is that the system size $R$ must be large enough to contain $c/u$ times 
the Larmor radius $r_g$ of a CR with energy $T_{\rm H}$;
$
r_g= 1.1 \left ( {T}/{\rm EeV} \right ) \left ( {B} /{\mu {\rm G}} \right ) ^{-1}\ {\rm kpc}
$.
CR reaching the Hillas limit at a relativistic shock have a Larmor radius comparable with the spatial extent $R$ of the shock.

The magnitude of $B$ is crucial.  At non-relativistic shocks in supernova remnants (SNR), CR drive an instability that amplifies the magnetic field (typically to 100s of $\mu{\rm G}$) far beyond that in the ambient medium (Bell 2004).  There are other means by which the magnetic field might be amplified (Giacalone \& Jokipii 2007) but these amplify the field over a large distance downstream of the shock and we do not consider them here since field amplification needs to take place close to a relativistic shock. 
Similar, presumably CR-driven, magnetic field amplification appears to happen at the termination shocks of the jets of radio galaxies where the 
hot-spot field is again typically 100s $\mu{\rm G}$ (eg Araudo et al 2016).  Magnetic field amplification is a general feature of optimal CR acceleration to high energy since the magnetic energy density upstream of the shock can be expected to be much smaller than the kinetic energy density $(\Gamma_{\rm sh} -1) \rho c^2$ produced by any event releasing sufficient energy to be capable of particle acceleration to high energy  (where $\rho $ is the density and $\Gamma_{\rm sh}$ the Lorentz factor of the bulk flow).

The magnetic field $B$ in the expression for $T_{\rm H}$ is therefore generally a turbulent magnetic field amplified by the CR.
This has two consequences.  First, the magnetic energy density cannot be larger than the energy density of the CR amplifying the field.
Second, the scale-size $s$ of the turbulent magnetic field cannot exceed the Larmor radius of the CR driving the turbulence
as discussed in Section 4. 
Magnetic field structured on a scale much smaller than the CR Larmor radius is ineffective in scattering the CR.
Hence, low energy CR with a small Larmor radius contribute little to the scattering of high energy CR.  The scattering of CR with a given energy is dominated by turbulent magnetic field generated by CR with that same energy.  And consequently, as discussed below, the  energy density  of the magnetic field scattering CR with a given energy cannot greatly exceed the energy density of CR with that energy.  
A further apparent consequence of strong turbulent magnetic field amplification 
is that any uniform field existing upstream of the shock, being comparatively small, 
has little effect on CR transport.
Shocks can then be treated as unmagnetized  in the
sense that magnetic field ordered on a large supra-Larmor spatial  scale is negligible.   
We will show in Section 5 that this last statement, while being true for CR at most energies, needs severe qualification.  In the case of perpendicular relativistic shocks, the uniform magnetic field turns out to be important in determining the maximum CR energy.
However, for the moment we will assume that CR transport is dominated by the turbulent disordered magnetic field and treat shocks as unmagnetised on the large scale.  Equivalently, this is the same as treating shocks as parallel rather than perpendicular since a large scale magnetic field parallel to the shock normal has no significant effect on CR transport.

The importance of the scale-size $s$ is seen in Lagage \& Cesarsky (1983a,b) and Drury (1983) who estimated a characteristic time for CR acceleration: 
$$
t_{\rm accel}= \frac{3}{u_{sh}^2} \left ( \frac{rD_u+r^2D_d}{r-1} \right )
\eqno(2)
$$
where $D_u$ and $D_d$ are the CR spatial diffusion coefficients upstream and downstream of the shock respectively, $u_{\rm sh}$ is the shock velocity, 
and $r$ is the compression at the shock (that is, $u_{\rm sh}/r$ is the downstream flow velocity relative to the shock). 
The maximum acceleration rate occurs if the diffusion coefficient is of the order of Bohm diffusion, 
$D_{\rm Bohm}=cr_g/3=T/3B$, since the CR scattering mean free path $\lambda$ cannot be smaller than the Larmor radius $r_g$.
Acceleration must take place in time $R/u_{\rm sh}$,
so the maximum CR energy is defined by $t_{\rm accel}=R/u_{\rm sh}$.
Assuming a fixed ratio for $\lambda/r_g$, and estimating that the magnetic field is compressed and increased by a factor $r_B$ at the shock, 
the maximum CR energy is
$
[ r_B/r][(r-1)/(r_B +1)]  (r_g / \lambda ) u_{\rm sh} BR
$
, where $1 \le r_{\rm B} \le r$ depending on the angle between the magnetic field and the shock normal.  
For strong ultra-relativistic and non-relativistic shocks
$r$ takes the values 3 (Kirk \& Duffy 1999) and 4 repectively.
For simplicity we make the approximation that $u_{\rm sh}=c$,  $r_{\rm B}=r=3$, 
and so define a `Lagage \& Cesarsky' (LC) limit as
$$
T_{\rm LC}= \left ( \frac {r_g } { \lambda } \right )   \frac { cBR}{2}\ .
\eqno{(3)}
$$
Observational evidence (Stage et al 2006, Uchiyama et al 2007) indicates that Bohm diffusion applies at non-relativistic shocks.
Furthermore, the diffusion has to be approximately Bohm for SNR to accelerate Galactic CR to the knee in the spectrum observed at a few PeV.

In later sections we will see that the maximum CR energy is low at relativistic shocks because the CR  mean free path is often much larger than the Larmor radius since the spatial scale-size of the magnetic field is much smaller than $r_g$.
If the magnetic field is randomly orientated on a scale-size $s$, 
each cell scatters the CR through a small angle $s/r_g$,
and the mean free path is
$$
 \lambda \approx \frac {r_g^2}{s} 
\eqno{(4)}
$$
 (see for example, Ostrowski \& Bednarz (2002), Kirk \& Reville (2010), Lemoine \& Pelletier (2010), Sironi, Spitkovsky \& Arons (2013) and Reville \& Bell (2014)).
$r_g=T_{\rm  LC}/cB$, so
$$
T_{\rm  LC} = cB \sqrt{\frac{sR}{2}} \ .
\eqno(5)
$$
The energy $T_{\rm LC}$ is smaller than the Hillas energy $T_{\rm H}$ by a factor $\sqrt{s/2R}$.
This means that if CR are to be accelerated to energies approaching the Hillas limit, as seems necessary to reach  $\sim100\  {\rm EeV}$, 
then the magnetic field responsible for scattering them in the shock environment must be generated by the high energy CR themselves.
If turbulence is driven by the non-resonant hybrid (NRH) instability (Bell 2004) or the resonant instability (Lerche 1967,
Kulsrud \& Pearce 1969)
 the scale-size $s$ is equal to the CR Larmor radius or smaller.
If the turbulence is driven by CR protons with 1 GeV energy in a magnetic field of 1 $\mu{\rm G}$ at a shock with $R=1\ {\rm kpc}$, then $s \sim r_g \sim 3\times 10^{10}{\rm m}$ and $\sqrt{sR}=3\times 10^{-5} R$.
Other longer wavelength instabilities (eg Drury \& Falle 1986, Bykov et al 2011) may play a role in scattering CR, but they are unlikely to close the large gap between the values of $s$ and $R$.
Hence we assume that we need only consider magnetic field amplification driven by the highest energy CR because magnetic field on smaller scales is unable to accelerate CR to high energy.
More will be said about this assumption below.

\section{ THE MAGNITUDE OF THE MAGNETIC FIELD}
 
In later sections we consider the effect of non-Bohm diffusion ($\lambda \gg r_g$),
but first we consider the limit on the maximum CR energy arising from the condition 
that the energy density of the magnetic field responsible for scattering CR is limited 
to the energy density of the CR at the same energy.
In this section we assume that $\lambda = r_g$.

The standard CR energy spectrum generated by strong non-relativistic shocks is proportional to $T^{-2}$.
This spreads the CR energy density uniformly in $\log (T)$ with the consequence that a significant fraction of the available shock energy is given to the highest energy particles.
In contrast, relativistic shocks produce steeper CR spectra $\propto T^{-\beta}$ where $\beta$ is larger than 2.
Kirk et al (2000) (see also Achterberg et al (2001)) derive $\beta =2.23$, and simulations by Sironi et al (2013) produce a spectrum steeper than this. 
A commonly observed synchrotron spectral index for radio galaxies is $\alpha \sim 0.75$, corresponding to $\beta \sim 2.5$, and 
powerful radio galaxies tend to have even  steeper spectra (Blundell, Rawlings \& Willott (1999), Miley \& De Breuck (2001)).
If CR are injected into the acceleration process at low suprathermal (GeV) energies, even a small steepening of the spectrum substantially reduces the CR energy density at EeV energies.  This is important because EeV CR are most strongly scattered by turbulence generated by the EeV CR themselves, and the scattering magnetic energy density is correspondingly small.
CR at GeV energies have a large energy density and generate a large magnetic field, but its maximum scale is that of the Larmor radius of GeV CR, which is too small to scatter EeV CR effectively.

If $\Gamma _{\rm sh}$ is the Lorentz factor of the shock, the energy density available for CR acceleration is $(\Gamma _{\rm sh} -1)  \rho c^2$.
We define the fraction of this energy given to CR with energy $T$ in a logarithmic energy range ($\Delta T = T$) to be
$$
\eta (T) = \eta _0 (T/T_{\rm inj})^{-(\beta-2) }
\eqno{(6)}
$$ 
where $T_{\rm inj}$ is the injection energy.  
Estimates of CR acceleration efficiency in the Galaxy are varied  (eg  Drury 1983, Blandford \& Eichler 1987)  but the efficiency must be high to provide the measured Galactic  CR density, and observations of supernova remnants indicate that the CR pressure is dynamically significant (Helder et al 2012).
We take $\eta _0 = 0.1$, equivalent to $\sim 30 \%$  of the available energy going to CR, and ignore the dependence of the CR energy density on $\beta $ for a given $\eta _0$.
In a similar vein of approximation we assume that the injection energy is relativistic and that $T_{\rm inj} \sim \Gamma _{\rm sh} {\rm GeV}$.
Given the approximate nature of our calculation we simplify our notation by treating shocks as relativistic and ignoring the difference between
$\Gamma _{\rm sh}-1$ and $\Gamma _{\rm sh}$.
In the generous assumption of energy equipartition between CR at energy $T$ and the magnetic field amplified by CR at energy $T$, the 
magnetic field effective in scattering CR with energy $T$ is $B_{\rm T} \sim ( \mu _0 \eta (T) \Gamma _{\rm sh} \rho c^2)^{1/2}$, which is equivalent to
$B_{\rm T} \sim 500 \Gamma _{\rm sh} ^{1/2}  (n_e/10^{-4} {\rm cm}^{-3})^{1/2}   (T/T_{\rm inj})^{-(\beta-2)/2}  \mu {\rm G}$, or
$$
B_{\rm T} \sim   8 \times  10^{-4.5(\beta -2.4)}  \Gamma _{\rm sh} ^{(\beta -1)/2}
 \left (\frac{ n_e} {10^{-4} {\rm cm}^{-3}} \right ) ^{1/2}
$$
$$ 
\hskip 3 cm \times
 \left (\frac {T}{\rm EeV} \right )^{-(\beta -2)/2}   \mu {\rm G}
\eqno(7)
$$
where $n_e$ is the thermal electron density, and we have assumed that $\rho = 1.2 n_e m_p$. 
We note that $B_{\rm T}$  is not the total magnetic field, but only the component responsible for scattering CR with energy $T$.   
The total magnetic field is larger and is dominated by the field amplified by low energy suprathermal CR.
The corresponding LC limit with Bohm diffusion  $cB_{\rm T}R/2$  ($\lambda \approx r_g$ in equation 3)  is 
$$
T_{\rm Bohm} \sim  \xi_{B}(\beta ) \Gamma _{\rm sh} ^{(\beta -1)/\beta }
 \left (\frac{ n_e} {10^{-4} {\rm cm}^{-3}} \right ) ^{1/\beta}
\left ( \frac { R}{\rm kpc} \right )^{2/\beta} {\rm eV}
\eqno(8)
$$
where $\xi _{B} (\beta)=10^{(9 \beta+22.7 )/\beta}$, 
giving
$\xi(2.23)=1.5 \times 10^{19}$ and $\xi(2.5)=1.2 \times 10^{18}$
(see Fig 1 and Table 1).
We have labeled this energy $T_{\rm Bohm}$ because of the assumption that $\lambda = r_g$.
$T_{\rm Bohm}$ is much less than the CR energy achievable in the total magnetic field, for example $cBR/2=230\  {\rm EeV}$ if $R= 1 \ {\rm kpc}$
 and $B=500\ {\rm \mu G}$,
but $T_{\rm Bohm}$ still exceeds 1 EeV.

The choice of $10^{-4} {\rm cm}^{-3}$ as the characteristic density may appear to be unnecessarily low, but a higher density would imply an uncomfortably large energy flux through the shock unless a smaller characteristic scale-size $R$ were chosen.
The characteristic parameters adopted in equation (8) are appropriate for jet termination shocks of radio galaxies.
Alternatively a larger density and smaller scale-size might be chosen.
However, as shown by Waxman (1995, 2001) and Blandford (2000), the choice of realistic parameters is limited by the energy flux through the shock.
Equation (8) can be recast in terms of the shock power, 
$P_s= \pi R^2 \Gamma _{\rm sh} \rho c^3$,
or equivalently 
$$
P_s = 1.5 \Gamma_{\rm sh} \left (\frac {n_e}{10^{-4}{\rm cm}^{-3}} \right )\left ( \frac {R}{\rm kpc} \right )^2 10^{40} {\rm W} \ .
\eqno(9)
$$ 
Assuming as above that a fraction $\eta (T) $ of this power is transferred to CR and thence
by equipartition to the magnetic field,
$$
T_{\rm Bohm} \sim \xi_B (\beta)  \Gamma _{\rm sh}^{-\frac{\beta-2}{\beta}}  \left (\frac{ P_s} {10^{40} {\rm W}} \right ) ^{1/\beta}
\ {\rm eV} \ .
\eqno(10)
$$
Even with the generous assumption of Bohm diffusion and equipartition magnetic fields, protons can be accelerated to a few EeV only if
energy is fed through the shock at a power exceeding $10^{40}{\rm W}$,
independent of the density and scale-size.
The power would need to be higher ($P_s \propto T_{\rm Bohm}^\beta $) for proton acceleration to 100 EeV.

\section{THE SCALE-SIZE OF THE MAGNETIC FIELD}

We have argued above that the turbulent magnetic field must dominate any ambient ordered magnetic field.
The maximum acceleration rate at shocks occurs if the CR diffusion coefficient is of the order of Bohm diffusion coefficient, 
$D_{\rm Bohm}=cr_g/3=T/3B$, for which the CR scattering mean free path $\lambda$ is approximately equal to the Larmor radius $r_g$.
We now show that the scattering mean free path $\lambda $ is in fact much larger than the CR Larmor radius $r_g$
as deduced from observations of termination shocks in radio galaxies (Araudo et al 2015, 2016).
This limits further the energy to which CR can be accelerated by relativistic shocks.
The magnetic field is frozen-in to the thermal plasma which behaves magnetohydrodynamically.
Magnetic field can only be generated on  a scale $s$ if the plasma into which it is frozen is moved a distance $s$ by the
${\bf j}_{\rm  CR} \times {\bf B}$ force through which the CR with current density ${\bf j}_{\rm  CR}$ acts on the background thermal plasma.  
An upper limit on the distance moved is therefore $s \sim j_{\rm  CR} B t^2/\Gamma _{\rm sh} \rho$
where $t$ is the time during which the force acts.

If the CR proton current $j_{\rm  CR}$ exists in a layer of thickness $L$ about the shock, then $t\sim L/c$.
The energy flux of CR with energy 
$T$ is $\sim \eta (T) \Gamma _{\rm sh} \rho c^3$, so $j_{\rm CR}\approx \eta (T)   \Gamma _{\rm sh} \rho c^3/T$ 
(Bell 2004)
giving
$$
s \sim  \eta (T) r_g \left ( \frac{L}{r_g}\right )^2
\eqno(11)
$$
where $r_g=T/cB$.
The ratio $L/r_g$ is expected to either remain constant or increase (become less Bohm-like) with increasing CR energy, but not decrease (become more Bohm-like).
Unless the CR spectral index is very steep ($\beta > 3$), $\eta (T)r_g$ ($\propto T^{-(\beta -3)}$) increases with energy.
Consequently $s$ increases with CR energy and, as discussed above, 
the scattering of CR at a particular energy is dominated by the turbulent magnetic field
generated by CR with that energy.
The scale-size of magnetic field generated by lower energy CR is too small to effectively scatter CR.

When considering the scattering of CR with the  highest energy ($T_{\rm max}$) we need only take account of the magnetic field
generated by these CR through their current density 
$j_{\rm CR}\approx \eta (T_{\rm max})   \Gamma _{\rm sh} \rho c^3/T_{\rm max}$.
The maximum characteristic time during which the ${\bf j}_{\rm CR} \times {\bf B}$ force acts is $R/c$,
so  the maximum displacement of the background plasma is
$j_{\rm CR} B R^2 /c^2 \Gamma _{\rm sh} \rho$, giving
$$
s_{\max} \sim \frac {\eta (T_{\rm max})    R^2}{r_g (T_{\max})}
\eqno(12)
$$
where $r_g(T_{\max})=T_{\rm max}/cB$.  
We assume that the magnetic field is randomly orientated in cells of size $s_{\max}$.
Equation (4) for the mean free path $\lambda $ gives
$ {r_g}/{\lambda } \sim \eta   R^2 / r_g^2 $.
In combination with equation (3), the maximum CR energy is then
$$
T_{\rm s} \sim   
\left ( \frac {\eta}{2} \right )^{1/3} cB_{\rm s}R
\eqno(13)
$$
where $ B _{\rm s} =[ \mu _0 \eta (T_{\rm s} )  \Gamma _{\rm sh} \rho c^2 ]^{1/2}$,
and we have labeled this energy $T_{\rm s}$ since it is derived by determining the displacement $s$ of the bulk fluid 
by the ${\bf j}_{\rm CR} \times {\bf B}$ force.
As in Section 3, the magnetic field $B _{\rm s} $ is assumed to be in energy equipartition with CR with energy $\sim T_{ s}$.
From equation (6) for $\eta (T)$, 
$$
T_s \sim  \xi _{s} (\beta)\ 
\Gamma _{\rm sh} ^{(5\beta -7)/(5 \beta -4)}
\hskip 10 cm
$$
$$
\hskip 1 cm
\times \left (\frac{n_e}{10^{-4} {\rm cm}^{-3}} \right ) ^{3/(5\beta -4)}
\left ( \frac{R}{\rm kpc} \right )  ^{6/(5\beta -4)}
\ {\rm eV}
\eqno(14)
$$
where $\xi _s (\beta)= 10^{(9 \beta +6.24 )/(\beta -0.8)} $.
giving $\xi_s(2.5)=8 \times 10^{16}$ and  $\xi_s(2.23)=3 \times 10^{18}$
(see Fig 1 and Table 1).

These equations show the combined effect of the relativistic steepening of the power law ($\beta > 2$) 
both for reducing the equipartition magnetic field, as in the previous section, and also for reducing the ${\bf j_{\rm CR}}\times {\bf B}$ force, 
reducing the distance by which the plasma can be moved, and thereby resulting in a diffusion coefficient  larger than Bohm ($\lambda > r_g$).  
For this reason, the energy $T_{\rm s}$ derived in this section is smaller than the energy $T_{\rm Bohm}$ given by equation (8).
The difference between $T_{\rm s}$ and $T_{\rm Bohm}$
is greater when the CR energy spectrum is steeper.
 
The estimate  $T_{\rm s}$ derived  in this section for the maximum CR energy  is based on the maximum turbulence 
scale-size that can be generated by the ${\bf j}_{CR} \times {\bf B}$ force operating  in the equipartition magnetic field (equation (7)). 
The magnetic field only reaches equipartition if the instability responsible for magnetic field amplification has sufficient time to grow.

\section{MAGNETIC FIELD AMPLIFICATION AT RELATIVISTIC SHOCKS}

As noted above, shocks only accelerate CR to high energy if the magnetic field is amplified beyond its ambient value,
in which case CR transport is dominated by small angle scattering and isotropic diffusion.
Acceleration to high energy is only possible if the CR currents are sufficient to drive the turbulence and generate
the amplified magnetic field in the time available.
This imposes an additional limitation on CR acceleration: that the linear growth time of the plasma instability responsible for amplifying the magnetic field
must be smaller than the time $t$ during which CR drive the instability. 

If the shock is parallel or unmagnetised and CR are free to propagate large distances away from the shock then $t \sim R/c$.
If the CR are scattered such that the mean free path $\lambda$ is smaller than $R$, then (equation (4)) $t \sim \lambda /c  \sim r_g^2/sc$ 
since $\lambda$ is the maximum spatial scale of the upstream CR precursor and also the distance over which momentum anisotropy decays downstream.
However, as discussed below in Section 5.1,  if the shock is perpendicular then CR currents exist only within a Larmor radius of the shock 
(Milosavljevic \& Nakar 2006) and
the relevant time is $t \sim r_{g0}/c$ where $r_{g0}=T/cB_{0\perp}$ is the CR Larmor radius in the uniform perpendicular magnetic field
$B_{0\perp}$.
If $\lambda < r_{g0}$ then $t\sim \lambda /c$ as with parallel shocks.
Overall, the time during which currents of CR with energy $T$ drive turbulence is
$$
t \sim \min \left ( \frac{R}{c},\frac{T}{c^2 B_{0\perp}},\frac {T^2}{s c^3 B_1^2}\right )
\eqno{(15)}
$$
where $B_1$ is the turbulent magnetic field,
and the distances determining the characteristic times in the brackets are $R$, $r_{g0}$ \& $\lambda$ ($=r_g^2/s$) respectively.  
 The time available for magnetic field amplification varies according to whether the shock is parallel or perpendicular,
and whether or not  the CR are strongly scattered.

It is well known (eg Kirk \& Duffy 1999) that ultra-relativistic shocks are much more likely to be
quasi-perpendicular than quasi-parallel with respect to the upstream ambient magnetic field.
CR moving along a field line (zero pitch angle) can only escape ahead of an oblique shock if the direction of the field line lies within an angle
$\theta = \sin ^{-1} (1/ \Gamma _{\rm sh} )$ of the shock normal.  
Upstream escape is further restricted if the CR pitch angle is non-zero and the CR spirals along the magnetic field.
Another reason why relativistic shocks behave quasi-perpendicularly is that the perpendicular component, unlike the parallel component, is increased by 
a factor $\Gamma _{\rm sh}$ when transforming from the upstream rest frame to the rest frame of the shock.
Additionally, compression of the perpendicular component as it traverses the shock further increases the perpendicularity of the magnetic field downstream of the shock.  Since relativistic shocks are characteristically perpendicular rather than parallel, we treat the perpendicular case first before returning to the case of parallel shocks.

The CR current around non-relativistic shocks is predominantly upstream of the shock as the CR diffuse ahead of the shock in a stationary precursor.
In contrast, the CR currents occur both upstream and downstream of relativistic shocks since strong anisotropies at the shock
take a while to decay as the CR are advected away downstream of the shock.
Figures 2 and 3 of Bell et al (2011) show that there can be strong downstream currents at perpendicular shocks even at shock velocities as low as $c/10$.
CR currents at high velocity perpendicular shocks are primarily diamagnetic anisotropies due to a deficiency of CR with gyrocentres upstream of the spatial point in question.
Downstream currents are also strong at fully relativistic parallel shocks in the downstream plasma frame since CR have to drift at a velocity greater than $c/3$ to re-enter the upstream plasma from downstream.
The NRH instability is excited by CR currents both upstream and downstream of the shock.
For simplicity, and in keeping with our level of approximation, we treat magnetic field amplification as though it occurs entirely downstream of the shock.
Note from Section 5.1.1 below, that the maximum CR energy depends on the uniform magnetic field  ${\bf B}_0$ at perpendicular shocks where
the CR current is strong downstream of the shock.
The appropriate magnitude of  ${\bf B}_0$ in the calculation is its compressed downstream value.

\subsection{Strong magnetic field amplification at perpendicular shocks}
Here we demonstrate that the plasma instability generating the turbulent magnetic field must be able to amplify the magnetic field by a large factor in 
 the time
$r_{g0}/c$ during which the plasma advects one Larmor radius downstream from the shock.  

Our model is as follows.  
A uniform perpendicular magnetic field exists upstream of the shock.  
The uniform field is compressed to a downstream magnitude $B_0$ as it passes through the shock.  
Because the magnetic field increases at the shock, the CR drift along the shock front with a current density $j_{CR} \sim \eta \Gamma _{\rm sh} \rho c^3/T$ distributed over a layer of thickness $r_{g0}=T/cB_0$  around the shock in the absence of scattering. 
${\bf j}_{\rm CR}$ is directed along the shock surface and perpendicular to ${\bf B}_0$.  
The ${\bf j}_{\rm CR} \times {\bf B}_0$ acts along the shock normal to slow down the background plasma flowing through the shock, but it also excites the NRH instability (Bell 2004) with {\bf k} vector aligned parallel to ${\bf B}_0$ as derived by Bell (2005).  
Riquelme \& Spitkovsky (2010) have investigated in depth the instability in this perpendicular geometry (${\bf j}_{\rm CR}$ perpendicular to ${\bf B}_0$),
referring to it as the  perpendicular current-driven instability (PCDI), using a 2D PIC code.
Using a 3D MHD code with an imposed uniform CR current, Matthews et al (2017) compare the perpendicular geometry with the parallel geometry  (${\bf j}_{\rm CR}$ parallel to ${\bf B}_0$ as in Bell (2004)). 
Riquelme \& Spitkovsky and Matthews et al show that the two geometries have closely similar linear growth rates (consistent with the dispersion relation derived as equation (4) of  Bell (2005)).
Matthews et al find closely similar parallel and perpendicular non-linear growth rates  (consistent with simulations of the parallel geometry by Beresnyak \& Li (2014)).
 Moreover, the instability converges to a similar non-linear spatial structure in both geometries.
In its 3D non-linear turbulent phase, when the magnetic field has been significantly amplified, the instability loses sense of the initial direction of ${\bf B}_0$ relative to  ${\bf j}_{\rm CR}$.
In both geometries the instability converges to a cavity/wall/loop structure qualitatively similar to the expanding loop version of the instability
analysed in section 2.2 of Bell (2005).  
Given the insensitivity to the orientation of ${\bf j}_{\rm CR}$ relative to ${\bf B}_0$ and the approximate nature of our calculations, we apply a common analysis to instability growth without distinguishing between parallel and perpendicular geometries of the NRH instability.
We consider the magnetic field to be the sum of the large-scale uniform field of magnitude $B_0$ and a turbulent field of magnitude $B_1$ generated by the NRH instability and growing to be much larger than $B_0$.

As in Section 4, we estimate the maximum distance $s$ that the 
${\bf j}_{\rm CR}\times {\bf B} _1$ force can displace the plasma in the available time, which we take to be $\sim r_{g0}/c$ in this section, 
and thereby estimate the scale-size of the turbulently generated magnetic field:
$s \sim j_{CR} B_1 (r_{g0}/c)^2 /\Gamma _{\rm sh} \rho$
where $j_{\rm CR} \sim  \eta (T) \Gamma_{\rm sh} \rho c^3/T$ and $r_{g0}=T/cB_0$, giving
$$
s \sim \eta (T)  r_{g0} \left ( \frac  {B_1} {B_0} \right ) \ .
\eqno(16)
$$
Since $\eta \ll 1$ for high energy CR, $s \ll r_{g0}$ unless the amplified magnetic field is much larger than the uniform field.

CR are accelerated at a relativistic quasi-perpendicular shock only if CR are strongly scattered within a distance of one Larmor radius downstream of the shock.  That is, the scattering mean free path immediately downstream of the shock 
must be less than a Larmor radius, $\lambda < r_{g0}$ (Lemoine \& Pelletier 2010, Reville \& Bell 2014). 
Since  $\lambda \sim r_g^2/s$ (equation (6)) where $r_g$ is the CR Larmor radius in the amplified field, 
the condition that $\lambda < r_{g0}$  imposes a requirement of strong magnetic field amplification, $B_1\gg B_0$:
$$
\frac {r_g ^2}{s} < r_{g0}
\hskip 0.35 cm
{\rm or \ equivalently \ } \hskip .25 cm
\left ( \frac {B_1}{B_0} \right ) ^2 > \frac{r_{g0}}{s}
\eqno(17)
$$
which shows that the scale-size $s$ of the turbulence can be smaller than the Larmor radius in the uniform field $B_0$
provided the amplified field $B_1$  is larger than $B_0$.
Combining relations (16) \& (17) gives a condition for successful CR acceleration:
$$
\frac{B_1}{B_0} > \eta ^{-1/3}
\eqno(18)
$$
Using equation (6) for $\eta (T)$, we find
$$
\frac{B_1}{B_0} > 2.2 \left ( \frac{T}{T_{\rm inj}} \right )^{(\beta -2)/3}
\eqno(19)
$$
The exponent $(\beta -2)/3$ is small for our representative values of $\beta $ (2.23 and 2.5),
but for acceleration to high energy, $T/ T_{\rm inj} \gg 1$,
significant magnetic field amplification is needed during the time $t \sim r_{g0}/c$ in which 
the plasma advects a distance of one Larmor radius $r_{g0}=T/cB_0$ in the uniform field $B_0$.
The CR current needs to be able to drive the amplification process for many instability growth times.

\subsubsection{ Magnetic field amplification and the maximum CR energy at relativistic perpendicular shocks}
Since the scale-size $s$ is much larger than the thermal Larmor radius, the field must be amplified by a MHD instability,
most probably the NRH instability as discussed in the previous section.
For any other MHD process to drive non-linear turbulence more effectively
and to enable CR acceleration to higher energy,  
it would have to make better use of the ${\bf j} _{\rm CR}\times {\bf B}$ force,
which appears unlikely from the simple dynamical argument in section 4.
Kinetic instabilities such as the Weibel instability operate on a scale which is too small to scatter UHECR.

The NRH instability has a maximum growth rate $\gamma _{\rm max}=0.5 j_{CR} \sqrt {\mu _0 /\Gamma_{\rm sh} \rho}$ 
(Bell 2004) where $j_{\rm CR} \sim  \eta  \Gamma _{\rm sh} \rho c^3/T$ as above.
The NRH instability has the advantage that it can amplify the magnetic field by orders of magnitude 
since it continues to grow rapidly in its non-linear phase ($\delta B/B \gg 1$).
As discussed by Bell et al (2013), an approximate condition for the time $t$ taken for amplification by a factor 10-100
in parallel geometry is given by  $\gamma_{\rm max} t \sim 5-10 $. 
As noted above, growth is very similar in parallel and perpendicular geometry.
Matthews et al (2017) show that the main difference is that in the perpendicular geometry the transition from linear growth to the slower non-linear growth occurs at a smaller amplitude, 
thereby marginally increasing the time needed to reach a given non-linear amplitude.

CR acceleration is only possible if the scattering mean free path $\lambda$ is smaller than
the CR Larmor radius $r_{g0}$, and we have argued in previous sections that scattering of CR with energy $T$
is dominated by turbulent magnetic field generated by CR with the same energy $T$.
If the magnetic field is amplified by the NRH instability, then the maximum CR energy $T_{{\rm NRH},\perp}$ for acceleration
at a perpendicular shock in magnetic field amplified by the NRH instability is determined by the 
condition $\gamma_{\rm max} t \sim 5-10 $
for $t =T_{{\rm NRH},\perp}/cB_0$ and 
$\gamma _{\rm max}=0.5 j_{CR} \sqrt {\mu _0 /\Gamma_{\rm sh} \rho}$ 
where $j_{\rm CR} \sim  \eta (T_{{\rm NRH},\perp})  \Gamma _{\rm sh} \rho c^3/T_{\max}$ .
Hence $T_{{\rm NRH},\perp}$ is determined by the equation
$$
\eta (T_{{\rm NRH},\perp}) \sim 10 \left ( \frac{B_{0\perp}^2/ \mu _0}{\Gamma _{\rm sh} \rho c^2} \right )^{1/2} \ .
\eqno(20)
$$
From equation (6) for $\eta (T)$, this maximum CR energy is

$$
T_{{\rm NRH},\perp}\sim
\xi _\perp (\beta) \ \Gamma _{\rm sh} ^{(2\beta-3)/(2 \beta -4)} 
\hskip 10 cm $$
$$
\times 
 \left (\frac{n_e}{10^{-4} {\rm cm}^{-3}} \right )^{1/(2 \beta -4)}
 \left (\frac{B_{0\perp}}{\mu {\rm G}} \right )^{-1/( \beta -2)}
\ {\rm eV}
\eqno(21)
$$
where $\xi _\perp (\beta) =10^{(9 \beta - 16.8 )/(\beta -2)}$, giving
 $\xi _\perp (2.5) =2.5\times 10^{11}$ and 
 $\xi _\perp (2.23) =1.6\times 10^{14}$
(see Fig 1 and Table 1).
This energy is labelled $T_{{\rm NRH},\perp}$ because it is limited by the growth of the NRH instability at a perpendicular shock.
CR beyond this energy are unable to generate turbulence before passing a Larmor radius $r_{g0}$ downstream of the shock.
Their scattering length $\lambda $ in given turbulence increases strongly in proportion to $T^2$, whereas their Larmor radius $r_{g0}$
increases only in proportion to $T$, so they are carried away downstream without returning to the shock for further acceleration.

\subsection{Perpendicular shocks with low magnetization}

This estimate (equation (21)) of the maximum CR energy falls far short of 100 EeV needed to explain the acceleration of UHECR. 
Equation (21) might be interpreted as an indication that large CR energies might be reached if the upstream magnetic field is very small.
However, the above analysis assumes that the CR Larmor radius in the shock-compressed upstream magnetic field is smaller than the shock radius $R$.
This condition only holds if $B_{0}$ exceeds a critical value $B_{\rm crit} =   (T_{{\rm NRH},\perp}/{\rm EeV}) (R/{\rm kpc})^{-1} \mu {\rm G}$.
If $B_{0} < B_{\rm crit}$, the Larmor radius is greater than the shock radius and the time available for magnetic field amplification
is $t\sim  R/c$ which is smaller than the value $r_{g0}/c$ assumed here.
At the limits of the perpendicular shock analysis, $B_{0}=B_{\rm crit}$, in which case
$$
B_{\rm crit}= \Gamma_{\rm sh} ^{(2\beta-3)/(2 \beta -2)}  
\left (\frac{n_e}{10^{-4} {\rm cm}^{-3}} \right )^{1/(2 \beta-2)}
\left (\frac{R}{\rm kpc} \right )^{-(\beta -2)/(\beta -1)}
\hskip 2 cm $$
$$
\hskip 2 cm \times
10 ^{-(9 \beta -19.2)/(\beta -1)}
\ {\mu {\rm G}}\ .
\eqno(22)
$$
$$
{\rm If\ } \beta=2.5,   \hskip 12 cm
$$
$$
B_{\rm crit}\approx   \Gamma _{\rm sh} ^{0.67} \left (\frac{n_e}{10^{-4} {\rm cm}^{-3}} \right )^{0.33} \left (\frac{R}{\rm kpc} \right )^{-0.33} 0.006 \ {\mu {\rm G}}
\  .
$$
$$
{\rm If\ } \beta=2.23, \hskip 12 cm
$$
$$
B_{\rm crit}\approx  \Gamma _{\rm sh} ^{0.59} \left (\frac{n_e}{10^{-4} {\rm cm}^{-3}} \right )^{0.41} \left (\frac{R}{\rm kpc} \right )^{-0.19} 0.2 \ {\mu {\rm G}}
\ .
$$
If $B_{0} \le B_{\rm crit}$,
the appropriate analysis is that of a parallel or unmagnetised shock since the uniform field $B_{0}$ 
is not large enough to confine the CR within a distance $R$ of the shock.
If $B_{0} \le B_{\rm crit}$ the maximum CR energy is that derived in the next section for parallel and unmagnetized shocks.

\subsubsection{The maximum CR energy at parallel or low-magnetization relativistic shocks}
If  $B_0<B_{\rm crit}$ or the shock is  parallel instead of quasi-perpendicular, the time available for magnetic field amplification is $t=R/c$ instead of $t=r_{g0}/c$.
The condition for successful CR acceleration is then  $(\eta  \Gamma _{\rm sh} \rho c^3/T ) (\mu _0/\Gamma_{\rm sh} \rho)^{1/2} (R/c) > 10  $,
where we have again assumed that $\gamma _{\rm max} t\sim 5-10$ is required for magnetic field amplification.
The maximum CR energy as derived from this relation is:
$$
T_{{\rm NRH},\parallel }\sim
\xi _ \parallel (\beta)\ 
\Gamma _{\rm sh} ^{(2\beta -3)/(2 \beta -2)}
\hskip 10 cm
$$
$$
\hskip 1 cm \times 
\left (\frac{n_e}{10^{-4} {\rm cm}^{-3}} \right )^{1/(2 \beta-2)}
\left (\frac{R}{\rm kpc} \right )^{1/(\beta -1)}
 \ {\rm eV}
\eqno(23)
$$
where $\xi _{\parallel} (\beta )= 10^{(9 \beta +1.1 )/(\beta -1)}$, giving
 $\xi _{\parallel} (2.5 )= 5 \times 10^{15}$ and
 $\xi _{\parallel} (2.23 )= 1.6 \times 10^{17}$
(see Fig 1 and Table 1).
This energy is labelled $T_{{\rm NRH},\parallel }$ because it is limited by the growth of the NRH instability at a parallel
 or unmagnetized ($B_0 < B_{\rm crit}$) shock.

Using equation (9) for the shock power $P_s$,
equation (23) can be rearranged as 
$$
T_{{\rm NRH},\parallel }\sim  
\xi _ \parallel (\beta)\ 
\Gamma _{\rm sh}^{(2\beta -4)/(2 \beta -2)}
\left (\frac{ P_s} {10^{40} {\rm W}} \right )^{1/(2\beta -2)}
\ {\rm eV}
\eqno(24)
$$
Hence in the case that  $B_0$ is very small and the shock is unmagnetized, the power requirement imposes
a stringent limit on the energy to which CR can be accelerated (see also Waxman 1995, 2001).

\section{ COMPARISON OF RELATIVISTIC AND NON-RELATIVISTIC SHOCKS}
CR acceleration by relativistic shocks is hampered by a number of factors:
\newline
(i) The CR energy spectrum is steeper than $T^{-2}$, so a smaller fraction of the shock energy is given to the highest energy CR
that drive the large-scale turbulence required to scatter the highest energy CR.
\newline
(ii) Relativistic shocks are predominantly quasi-perpendicular, so CR have to be strongly scattered within one Larmor radius
if they are to return to the shock from downstream.
\newline
(iii) Because of (ii), the ${\bf j}_{\rm CR} \times {\bf B}$ force has insufficient time to move  the bulk thermal plasma through one CR Larmor radius,
so the scale-size $s$ of the amplified magnetic field is small; it can only scatter the CR effectively if it is amplified to a magnitude much
larger than that of the uniform field $B_0$.
There is little time for CR-driven plasma instabilities to amplify the magnetic field.

In contrast, in the case of non-relativistic shocks:
\newline
(i)  The standard CR energy spectrum, $T^{-2}$ places equal energy densities in each logarithmic range of energy.
\newline
(ii)  Non-relativistic shocks are predominantly quasi-parallel rather than perpendicular; CR are able to diffuse far upstream, amplify the 
magnetic field before it is overtaken by the shock, and produce strong CR scattering both upstream and downstream of the shock.
\newline
(iii) The magnetic field has time to grow on Larmor scales and produce Bohm-like diffusion as indicated by observation
(Stage et al 2006, Uchiyama et al 2007).
\newline
(iv) The magnetic energy density has time to reach large values upstream of the shock on all scales.

Non-relativistic shocks have advantages over ultra-relativistic shocks for particle acceleration, 
but the difficulty with non-relativistic shocks is that the accelerating electric field ($\propto u_{\rm sh}B$) 
is reduced by the ratio of the shock velocity $u_{\rm sh}$ to the
CR velocity $c$.  
Also, the CR current is reduced by $\sim u_{\rm sh}/c$, thus reducing instability growth rates  (Bell et al 2013).
Shocks with $u_{\rm sh} \ll c$, as in Sedov-phase SNR, do not accelerate CR to high energies (Bell 2015).

Consequently, it appears that if shocks are to accelerate UHECR they probably must have velocities less than $c$ by a factor of a few, 
but not by a factor very much larger than this.
An important question is just how much less than $c$ the shock velocity must be for effective particle acceleration.  
Araudo et al (2015, 2016) show from observations that extragalactic jet termination shocks with velocities around $c/3$ are poor accelerators.
Similarly, the highest velocity SNR ($u_{\rm sh} \sim c/10 - c/5$) have steep synchrotron spectra indicating ineffective CR acceleration (Bell et al, 2011).
These observations suggest that $u_{\rm sh} \sim c/5$ might be optimal for CR acceleration,
but these cases might be special since in both cases the shocks may be perpendicular.
In the case of extragalactic jets, the magnetic field in the jet might be predominantly toroidal.
Similarly, young fast SNR may expand into a magnetic field
in the form of a Parker spiral.
Mildly relativistic shocks ($u_{\rm sh} \sim c/5-c/2$) expanding into a randomly orientated magnetic field are equally likely to be quasi-perpendicular or quasi-parallel.  
From theory and simulations it is known that the CR energy spectrum at mildly relativistic shocks 
can be either flattened or steepened relative to $T^{-2}$ with a spectral index 
dependent on the obliqueness of the shock  (Kirk \& Heavens 1989, Bell et al, 2011).
It is possible that mildly relativistic parallel, but not perpendicular, shocks might be good UHECR accelerators.

\section{THE ACCELERATION OF HEAVY NUCLEI}
Thus far throughout the paper we have considered only the acceleration of protons. 
If an upstream plasma consisted entirely of  fully ionized atoms with charge $Ze$ and mass $Zm_p$  ($A=Z$), with ion number density $n_i$ in the shock rest frame, it would behave exactly the same as a plasma consisting entirely of fully ionized  hydrogen with number density $Zn_i$ except that the energies of individual nuclei would be $Z$ times larger.
The situation can be more complicated when the plasma consists of a mix of species with different $A/Z$ at different energies
as observed by the Auger array (Molina Bueno et al 2015).
The plasma turbulence will be driven by CR with the largest current density, which can be expected to be protons in most cases.
The amplified magnetic field would have an amplitude and scalelength determined by protons, 
their injection efficiency, their spectra, their spatial distribution, and their maximum energy.
It is not clear how this affects the acceleration of nuclei with a different $A/Z$,
but in general one can still expect heavier nuclei to be accelerated to higher energy than protons.

\section{ CONCLUSIONS}
We have shown that the maximum energy of CR accelerated by relativistic  shocks falls far short of the Hillas energy.
In Section 3 we derived an expression for the maximum CR energy $T_{\rm Bohm}$ as reduced by the steepness of the CR spectrum but assuming Bohm diffusion.
In Section 4 we showed that the maximum CR energy is further reduced to $T_{\rm s}$ when the departure from Bohm diffusion due to the small spatial scale-size $s$ of the magnetic field  is allowed for.
In Section 5 we derived an expression for the maximum CR energy $T_{{\rm NRH},\perp}$ \& $T_{{\rm NRH},\parallel}$ when the condition is applied that there must be time for the magnetic field to be amplified by the NRH instability.
The maximum CR energy is particularly reduced in the case of quasi-perpendicular shocks characteristic of relativistic flows.
$T_{\rm Bohm}$ and $T_{\rm s}$ are upper limits to the maximum CR energy since both are derived for parallel shocks and 
neither allows for the additional limitations imposed by the time needed to amplify the magnetic field.
$T_{{\rm NRH},\perp }$ may offer a reasonable estimate of the maximum CR energy,
but other factors may intervene to further limit acceleration and make this also an upper limit.
As shown in figure 1, $T_{{\rm NRH},\perp}$  depends strongly on the spectral index $\beta$ of the CR energy distribution
since the NRH instability is driven by the electric current carried by high energy CR.

Our overall conclusion is that if UHECR are accelerated by shocks then the shocks are probably non-relativistic, or possibly mildly relativistic, 
but not ultra-relativistic.  
The strongest limit on the maximum CR energy arises from the need to amplify a turbulent magnetic field within one Larmor radius of the shock.
The maximum CR energy is larger if the shock is parallel, but parallel relativistic shocks are a special and unlikely case.
The maximum CR energy is also potentially larger if the upstream plasma is unmagnetized such that the CR Larmor radius is larger than the size $R$ of the shock, but in this case the need for a large shock power limits the CR energy.

\section {ACKNOWLEDGEMENTS}
We are grateful to many people who have commented on this work, especially participants of the 
workshop entitled 'Beyond a PeV: Particle acceleration to extreme energies in cosmic sources' organised by Martin Lemoine and Guy Pelletier at the Institut d'Astrophysique de Paris, Sept 2016.
This research was supported by the UK
Science and Technology Facilities Council under grant No. ST/K00106X/1 and ST/N000919/1.


\section{REFERENCES}

Achterberg A, Gallant YA, Kirk, JG, Guthmann AW, MNRAS 328, 393 (2001)
\newline
Araudo AT, Bell AR, Blundell KM, ApJ 806, 304 (2015)
\newline
Araudo AT, Bell AR, Crilly A, Blundell KM, MNRAS 460, 3554 (2016)
\newline
Axford WI, Leer E, Skadron G,  Proc 15th Int. Cosmic Ray Conf. 11, 132 (1977)
\newline
Bell AR,  MNRAS 182, 147 (1978)
\newline
Bell AR, MNRAS 353, 550 (2004)
\newline
Bell AR, MNRAS 358, 181 (2005)
\newline
Bell AR, MNRAS 447, 2224 (2015)
\newline
Bell AR,  Schure KM, Reville B, MNRAS 418, 1208 (2011)
\newline
Bell AR, Schure KM, Reville B, Giacinti G, MNRAS 431, 415 (2013)
\newline
Beresnyak A, LI H, ApJL 788, 107 (2014)
\newline
Blandford RD,  Eichler D,  Phys Rep 154,1 (1987)
\newline
Blandford RD,  Ostriker JP,  ApJ 221, L29 (1978)
\newline
Blandford RD, Phys Scripta   T85, 191 (2000)
\newline
Blundell KM, Rawlings S, Willott CJ,  AJ 117, 677 (1999).
\newline
Bykov AM, Osipov SM, Ellison DC, MNRAS 410, 39 (2011)
\newline
Drury LO'C, Rep Prog Phys, 46, 973 (1983)
\newline
Drury LO'C, Falle SAEG,  MNRAS 222, 353 (1986)
\newline
Helder EA, Vink J, Bykov AM, Ohira Y, Raymond JC, Terrier R, Spce Sci Rev 173, 369 (2012)
\newline
Hillas AM ARA\&A 22, 425 (1984)
\newline
Giacalone J, Jokipii JR ApJL 663, L41 (2007)
\newline
Kirk JG, Duffy P JPhysG 8, R164 (1999)
\newline
Kirk JG, Guthmann AW, Gallant YA \& Achterberg A, ApJ 543, 235 (2000)
\newline
Kirk JG, Heavens AF,  MNRAS 239, 995 (1989)
\newline
Kirk JG,  Reville B, ApJL 710, L16 (2010)
\newline
Krymskii GF,  Sov Phys Dokl, 23, 327 (1977)
\newline
Kulsrud R, Pearce WP, ApJ 156, 445 (1969)
\newline
Lagage PO, Cesarsky CJ, A\&A 118, 223 (1983a)
\newline
Lagage PO, Cesarsky CJ, ApJ 125 249 (1983b)
\newline
Lemoine M, Pelettier G, MNRAS 402, 321 (2010)
\newline
Lerche I,  ApJ 147, 689 (1967)
\newline
Matthews JH, Bell AR, Blundell KM, Araudo AT, MNRAS 469, 1849 (2017)
\newline
Miley G, De Breuck C.,  2008, AAR, 15, 67
\newline
Milosavljevic M, Nakar E, ApJ 651, 979 (2006)
\newline
Molina Bueno L, Pierre Auger consortium, AN 785, 336 (2015)
\newline
Ostrowski M, Bednarz J, A\&A, 394, 1141 (2002)
\newline
Reville B, Bell AR, MNRAS 439, 2050 (2014)
\newline
Riquelme MA, Spitkovsky A, ApJ 717, 1054 (2010)
\newline
Sironi L, A Spitkovsky A, Arons J, ApJ 771, 54 (2013)
\newline
Stage MD, Allen GE, Houck JC, Davis JE, Nat Phys 2, 614 (2006)
\newline
Uchiyama Y, Aharonian FA, Tanaka T, Takahashi A, Maeda Y, Nat 449, 576 (2007)
\newline
Waxman E, Phys Rev Lett 75, 386 (1995)
\newline
Waxman E, in {\it Physics and Astrophysics of Ultra-High-Energy Cosmic Rays}, 
 ed Lemoine \& Sigl publ Springer (Berlin),
Lecture Notes in Physics 576, 122 (2001)

\end{document}